\documentclass[conference]{IEEEtran}
\IEEEoverridecommandlockouts
\usepackage{cite}
\usepackage{amsmath,amssymb,amsfonts}
\usepackage{algorithmic}
\usepackage{graphicx}
\usepackage[utf8]{inputenc}
\usepackage{textcomp}
\usepackage{xcolor}
\usepackage{url}
\usepackage{comment}

\def\BibTeX{{\rm B\kern-.05em{\sc i\kern-.025em b}\kern-.08em
    T\kern-.1667em\lower.7ex\hbox{E}\kern-.125emX}}
\begin{document}

\title{How often should I access my \\ online  social networks?
}

\author{
\IEEEauthorblockN{Eduardo Hargreaves}
\IEEEauthorblockA{Department of Computer Science \\
Federal University of Rio de Janeiro\\
Rio de Janeiro, RJ, Brazil\\
eduardo@hargreaves.tech}
\and
\IEEEauthorblockN{ Daniel Sadoc Menasché} 
\IEEEauthorblockA{Department of Computer Science \\
Federal University of Rio de Janeiro\\
Rio de Janeiro, RJ, Brazil\\
sadoc@dcc.ufrj.br}
\and
\IEEEauthorblockN{ Giovanni Neglia}
\IEEEauthorblockA{
Inria, Université C\^ote D'Azur \\
Sophia Antipolis, France\\
giovanni.neglia@inria.fr}
}

\maketitle

\begin{abstract}
Users of online social networks are faced with a conundrum of trying to be always informed without having enough time or attention budget  to do so. The  retention of users 
on online social networks has important implications, encompassing economic, psychological  and infrastructure aspects.  In this paper, we pose the following question: what is the optimal rate at which users should  access a social network? To answer this question, we propose an analytical model  to determine  the value of an access (VoA) to the social network.   In the simple setting considered in this paper, VoA is defined as the  chance of a user accessing the network and obtaining new content. Clearly, VoA depends 
 on the rate at which sources generate content and on the filtering imposed by the social network.  Then, we pose an optimization problem wherein the utility of users grows with respect to   VoA but is penalized by costs incurred to access the network. Using the proposed framework, we provide insights on the optimal access rate. Our results are parameterized using Facebook data, indicating the predictive power of the  approach.
\end{abstract}

\begin{IEEEkeywords}
Facebook, model, sampling, online social networks, age of information, value of access, News Feed, timelines
\end{IEEEkeywords}

\section{Introduction}
\label{sec:intro}

\paragraph*{Motivation} Users of online social networks are faced with a conundrum of trying to be always informed without having enough time or attention budget  to do so.  The latest version of iOS (12) and beta versions of Android send reports to users about the time they spend on apps installed on smartphones, and  allow users to limit their screen time~\cite{IOSScreenTime}. The time spent in social networks has important implications, which go beyond the social field, encompassing economic, psychological and network infrastructure aspects. From the economic point of view, in a world where ``a wealth of information creates a poverty of attention"~\cite{Simon}, attention becomes the scarce resource in the \emph{attention economy}~\cite{ShapiroVarian,crawford2015world}.

\paragraph*{Challenges}
On social networking platforms, the dispute for attention occurs through the counting of clicks and the time spent viewing publications. Users of social network platforms typically   consume content created by the sources they  follow. 
If users do not access online social network frequently, important posts might be missed. On the other hand, if users access too often, the chance that there will be no new relevant content created in the interval between accesses is high. Ultimately,  frequent accesses to social networks can be a source of challenges for  users (e.g., reflecting addictions or consuming resources unnecessarily) and for social network managers (e.g., lack of new information can lead to a drop in the quality of the experience of users impairing engagement with the platform).  Thus, determining the optimal frequency with which users should access a social network, taking into account the multiple and complex factors involved, is an important open problem.

\paragraph*{Gap on prior art} 
Most social media platforms get their revenue by displaying ads to their users. For this reason, the literature on user retention in social networks focuses primarily on the effect of  retention on profit, leveraging  sponsored advertisements and publications~\cite {jiang2013optimally}. In essence, it is assumed that the more frequently  a user accesses the network, the more likely he is to consume a sponsored product~\cite{Facebook_ads, Twitter_ads}. Users incur  monetary and attention-related costs to access social networking platforms, and if they access the network too often they end up either receiving repeated content  or content that does not necessarily suit their needs. \emph{We are unaware  of  previous work that  analyzed the optimal frequency with which users should access their social networks, considering the intrinsic value associated with each  access.}

\paragraph*{Methodology and goals} In this paper, we consider the problem of determining the optimal rate of access to a social network. To this purpose, we propose an analytical model that allows us to determine, depending on the rate at which sources generate content, the chance of a user accessing the network and obtaining new content. The proposed model yields a closed form expression for the value of an access (VoA), which we use as basis for an optimization problem.  In the optimization problem, the utility of each user increases with respect to VoA, and decreases with respect to the cost associated to an  access to the social network. The solution to the  optimization problem  provides {insights} on the optimal access rate. Our results are parameterized using Facebook data, which indicate the predictive power of the proposed approach.

\paragraph*{Contributions} In summary, our main contributions are listed below.


\begin{itemize}
    \item \textbf{New metric:}  We develop a new metric, referred to as \emph {value of access}, or VoA, to measure the amount of new information transmitted by a social network  platform per access.  The metric is evaluated using data collected from Facebook.
\item  \textbf{Analytical model:} We present an analytical model to estimate the expected number of new impressions for each view of a \emph{timeline}. The model produces simple closed form expressions for the metric of interest.
\item  \textbf{Optimal access frequency:} We pose an optimization problem that produces as  result the optimal frequency of access, balancing between the cost per access and the utility derived from it. Using the model, we analytically indicate  properties of the optimal solution.
\end{itemize}


\paragraph*{Paper structure} The remainder of the paper is organized as follows. In Section~\ref{sec:model} we introduce the analytical model that is validated in Section ~\ref{sec:validationger}. The problem of the optimal frequency of access  is posed in Section~\ref{sec:control}, and numerical examples are reported in Section~\ref{sec:numerical}. Related work and a  discussion of the practical implications of the work  follow in Sections~\ref{sec:related} and~\ref{sec:discussion} and Section~\ref{sec:conclusion} concludes.
\section{Model}
\label{sec:model}

Next, we describe the proposed model.  We consider sources that create posts that appear in  users timelines.  A user \emph{access}, also referred to as \emph{view}, is the act of refreshing a timeline.  A \emph{timeline}, in turn,  is a sorted list of posts presented to a user.  Every time a user accesses the network, a  timeline is prepared.  The timeline may contain posts that have already been presented to the user before.  Nonetheless, there are no repeated posts in a given timeline.  An \emph{impression} is a post appearing in a timeline of a user.

Let $\mathcal{I}$ and $\mathcal{J}$ be a set of users and sources, respectively.  
Let $i \in \mathcal{I}$ be a user following a  set of sources $\mathcal{J}_i$. The social network platform chooses the publications from $\mathcal{J}_i$ to be  displayed  in  user's $i$ {timeline}.  
Except otherwise noted, we focus on a given user $i$, and drop subscript from $\mathcal{J}_i$  whenever it is clear from context.

We assume that the set of sources followed by $i$ creates publications following a Poisson process with an aggregate rate of $\lambda$ publications per unit time.
A {timeline} of size $K$ contains  at most $K$ publications, i.e., the \emph{timeline size} is the maximum number of elements in a timeline.

Let $T$ be a random variable  characterizing the interval between accesses of a given user to the social network. The average interval between views is given by $\overline{T}$. In addition, we denote by $f_T(t)$ the probability density function of $T$. We remove the subscript when clear from the context, denoting the density simply as $f(t)$.

The \emph{access rate} (or view rate) is the rate at which a user accesses the network. 
We denote by $\mu$ be the user access rate, i.e., the frequency with which the user of interest accesses the social network.
The access rate (or view rate)  is
our main quantity of interest.


We denote by $\rho$  the average number of posts created between two user accesses. Then, we have
\begin{equation}
    \mu = \frac{1}{\overline{T}}, \qquad
    \rho = \frac{\lambda}{\mu}.
\end{equation}


\subsection{Value of Access (VoA) and utility}

At every access, a user  receives a timeline possibly containing novel impressions. A \emph{novel impression} is an impression corresponding to a post that has not yet been presented to that user,  i.e., an impression is considered \emph{new}  if it is included in the timeline of that user  for its first time. 
A \emph{repeated impression} is an impression corresponding to a post that has already been presented to that user. 

The \emph{value of access (VoA)} is the expected number of novel impressions in a timeline.
We assume that  the number of new impressions serves as a proxy for the amount of valuable information received, and we interchangeably refer to the VoA and to the expected number of new impressions per access as synonyms.

Let $V$ be the random variable denoting the  number of novel impressions in a timeline, and $\overline{V}$ its expected value, i.e., \emph{VoA}~$= \overline{V}$. 
In addition, let $A$ be
the random variable characterizing the number of publications created by the sources in the interval between two views. 
Table~\ref{tab: notation} together with   Appendix~\ref{sec:termino} summarize the notation used throughout this work.

The \emph{access cost} is the cost of accessing a social network, e.g., measured in monetary units, but reflecting the attention span dedicated to the access.  The \emph{utility of an access} is the value of the access minus the associated costs.


The analytical  model that follows produces the expected number of new impressions per view, $\overline{V}$,  and the utility per time unit, 
as a function of
\begin{itemize}
    \item  the size of the timeline ($K$),
    \item the publication creation rate ($\lambda$) and 
    \item the  rate at which the user accesses the social network and refreshes the timeline ($\mu$).
    \end{itemize}

\begin{table}[h!]
   \centering
   \caption{Table of notation (accesses and views are synonyms).}
   \begin{tabular}{l|l}
         variable & description \\
        \hline
         $\bar{V}$ & Value of Access (VoA); in this work, we assume equal to \\
         & number of \emph{new} impressions per access with fixed timeline size\\
         $\bar{V}^{(P)}$ & VoA assuming Poisson timeline size  \\
         & number of \emph{new} impressions per access \\
         $K$ & size of timeline (number of impressions per access) \\
         $\lambda $ &  rate of creation of publications (publications per unit time) \\
         $A$ & number of publications created between accesses \\
         $\mu $ & average access rate \\
         $1/\mu$ & average interval between accesses \\
        $\rho$ & average amount of new publications per access
  \end{tabular} 
   \label{tab: notation}
\end{table}

\subsection{Assumptions}

\label{sec:assumptions}


The analytical model is based on the following three assumptions:

\textbf{(A1) Content is generated according to a Poisson process} and the aggregate rate of content generation is $\lambda$.

\textbf{(A2) The timeline displays as many new content as possible. } If there is at least one new post that has not yet been displayed, and the user has space in the timeline to include such impression, i.e., if the number of newly created posts is smaller than or equal to $K$,  the content will be displayed  generating a new impression. Note that the model is agnostic with respect to the order at which the contents are displayed in the {timeline}

\textbf{(A3) Content already displayed will  be repeated if there are not enough novelties.} The  motivation for repeating impressions across timelines  is to fill the {timeline} if there are no new impressions available.


\subsection{Quantifying the value of access}

To compute the VoA, we first condition on the  time between two accesses. Let $\tau $ be the time between accesses (fixed and given).  
In this case, given the assumptions of the model, the number of new impressions per access is given by a truncated Poisson distribution,
Note that  a user can receive a maximum of $ K $ new impressions per refresh of his  {timeline},

\begin{equation}
V =\left\{    
\begin{array}{ll}
    A, & \textrm{if } A < K,  \\
    K, & \textrm{otherwise.} 
\end{array} \right.
\label{eq:cond_blocking}
\end{equation}

where $A$ is the number of publications created during $\tau$, 

\begin{equation}
    P(A=a|\tau) = \frac{e^{-\lambda \tau} (\lambda \tau)^a}{a!}, \label{eq:poiss}
\end{equation}

Hence,

\begin{equation}
    P(V=i|\tau) = \left\{ \begin{array}{ll}
         P(A<i|\tau), & \textrm{ if } i < K  \\
         P(A\geq i|\tau), & \textrm{ if } i = K \\
         0, & \textrm{ if } i > K  
    \end{array}\right.
    \label{eq:cases}
\end{equation}

The three conditions above correspond to situations in which $ (i) $ the generated content is insufficient to fill the timeline and $ (ii) $ the timeline is filled exclusively with new content. The third and last condition $(iii)$ captures the fact that the timeline has space for up to $K$ impressions.

It follows from~\eqref{eq:cases} that $E(V|\tau)$ is given by
\begin{align}
    E(V|\tau)&= \left(\sum_{i=0}^{K} i P(A=i|\tau) \right)+ K P(A > K |\tau).
    \label{eq:initial}
\end{align}

As $ P (A> K |\tau) = 1-P (A \leq K |\tau) $, we have:

\begin{align}
    E(V|\tau)&=\left(\sum_{i=0}^K i e^{-\lambda \tau} \frac{(\lambda \tau)^i}{i!}\right) + K \left(1- \sum_{i=0}^K e^{-\lambda \tau} \frac{(\lambda \tau)^i}{i!}\right)\\
          &=K- \sum_{i=0}^K (K-i) e^{-\lambda \tau} \frac{(\lambda \tau)^i}{i!} \label{eq:exprndet}
\end{align}

Finally, by applying the law of total probability,

\begin{align}
    \overline{V}&=\int_{\tau=0}^{\infty} E(V|\tau) f(\tau) d\tau.
        \label{eq:NGivenTau}
\end{align}

Next, we consider special cases of the expression above, as a function of the distribution of time between users' accesses.

\subsection{Special case I: deterministic time between accesses}

Let $\kappa$ be the deterministic time between samples.
Then,  

\begin{align}
    f(\tau)& = \delta_{\kappa}(\tau)  
\end{align}
where $\delta_{\kappa}(\tau)$ is a Delta Dirac function, with all mass at point $\kappa$. 

Note that in this case the number of posts created between two accesses is distributed according to a Poisson variable with mean  $\kappa \lambda$.
Then,
\begin{equation}
    \overline{V} =\int_{\tau=0}^{\infty} E(V|\tau) f(\tau) d\tau = E(V|\tau)
\label{eq:fixedtau}
\end{equation}
and \eqref{eq:fixedtau} is given by~\eqref{eq:exprndet}.
\subsection{Special case II: exponentially distributed interval between accesses}

Next, we consider an exponentially distributed  time between accesses.
Let $T$ be an exponentially distributed random variable with mean $ 1 / \mu$, i.e.,
$ f (\tau) = \mu e^{-\mu \tau}$. 
Substituting $ f (\tau) $ into 
\eqref{eq:NGivenTau},\begin{equation}
   \overline{V}=\int_{\tau=0}^\infty \left( K- \sum_{i=0}^K (K-i) e^{-\lambda \tau} \frac{(\lambda \tau)^i}{i!}\right)   \mu e^{-\mu \tau} d\tau. 
 \end{equation}

 After algebraic manipulation (see Appendix ~\ref{sec:simiplificaen}),  we get  a compact closed form expression for $\overline{V}$,
  
  \begin{align}
\overline{V}= \frac{\lambda}{\mu} \left(1-\left(\frac{\lambda}{\lambda+\mu} \right)^{K} \right).
\label{eq:NewImpressions}
\end{align}

\paragraph{Probabilistic interpretation}
As the interval between publications and the interval between accesses are exponentially distributed with rates equal to $\lambda $ and $\mu$, the ratio $\lambda / (\lambda +\mu) $ is the probability that  a new publication occurs before the user accesses the  social media~\cite{Durrett2006}. Then, due to the lack of memory of the exponential distribution, $ (\lambda / (\lambda +\mu))^K $ is the probability that at least $ K $ publications have occurred between two accesses.   In this case, such publications are sufficient to fill the timeline with novelties.  


Let $ p = 1 - (\lambda / (\lambda +\mu))^K $ be the probability that the user  accesses the system before the arrival of $ K $ consecutive  new publications. When $ p \approx 1$,
the user  will receive, on average, $\lambda /\mu $ new impressions per view.     

\paragraph{Limiting probabilities} 
The following limits apply:

\textbf{Size of the timeline} $\lim_{K\rightarrow\infty} \overline{V} =\lambda /\mu $. In this scenario, the user can read all the novelties created by the sources between two accesses, and the average number of novelties per access is equal to $\lambda /\mu$;

\textbf{Content generation} $\lim_{\lambda\rightarrow\infty}\overline{V} = K $ and $\lim_{\lambda\rightarrow 0} \overline{V} = 0 $ In this scenario, sources generate content at high (resp., low) rate, and the VoA is maximum (resp., minimum), and equal to $ K $ (resp., 0);

\textbf{Access frequency} $\lim_{\mu\rightarrow 0} \overline{V} = K $ and $\lim_{\mu\rightarrow \infty}\overline{V} = 0 $ In this scenario, users access the system with low  (resp., high) frequency, and the number of news per view is maximum (resp., minimum), and equal to $ K $ (resp., 0).


In the upcoming sections, we validate the proposed model using data collected from Facebook during the Brazilian elections in 2018.

\section{Experimental Results}
\label{sec:validationger}
\begin{figure*}
    \centering
\begin{tabular}{cc}
     \includegraphics[width=0.5\textwidth]{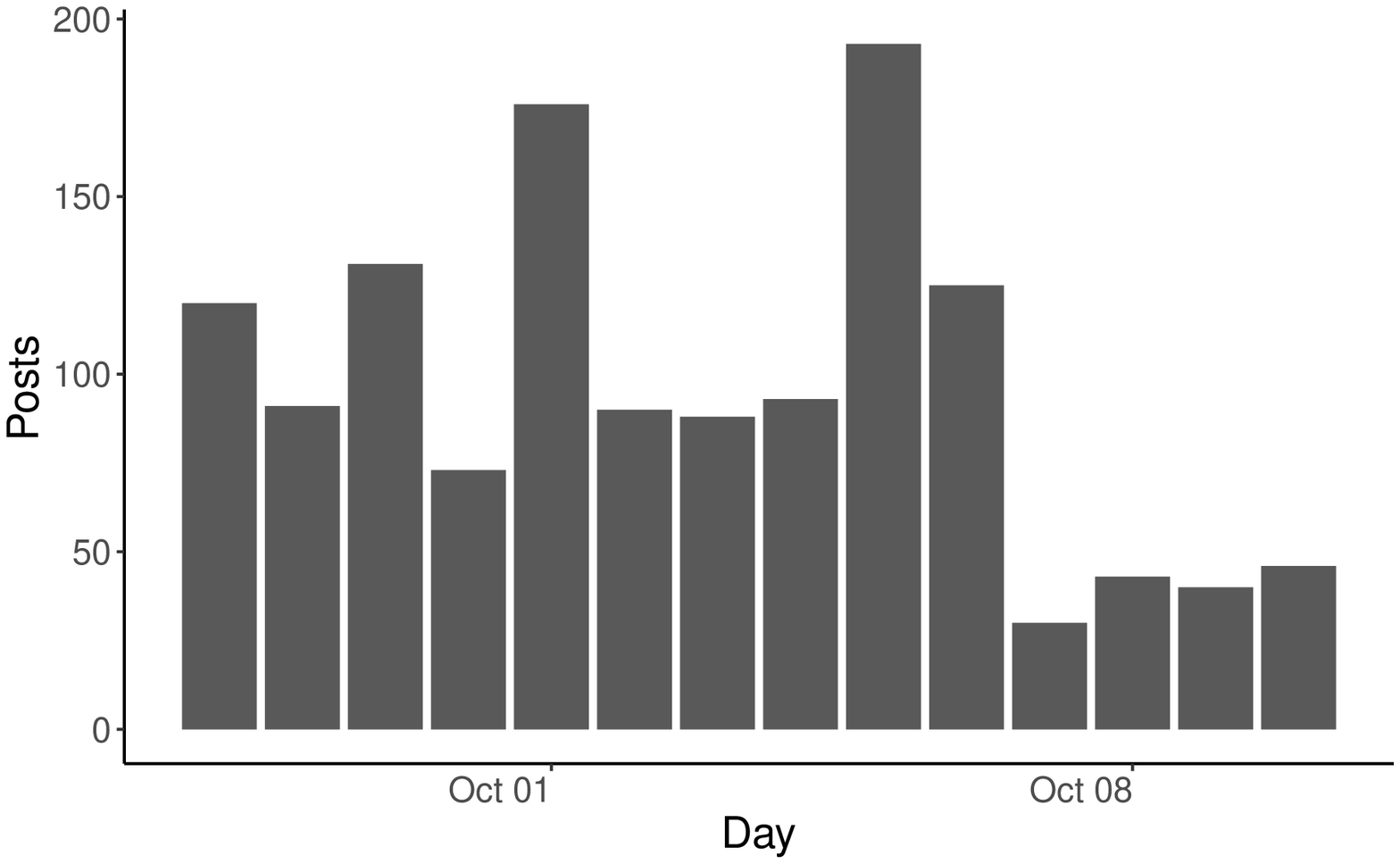} &      \includegraphics[width=0.5\textwidth]{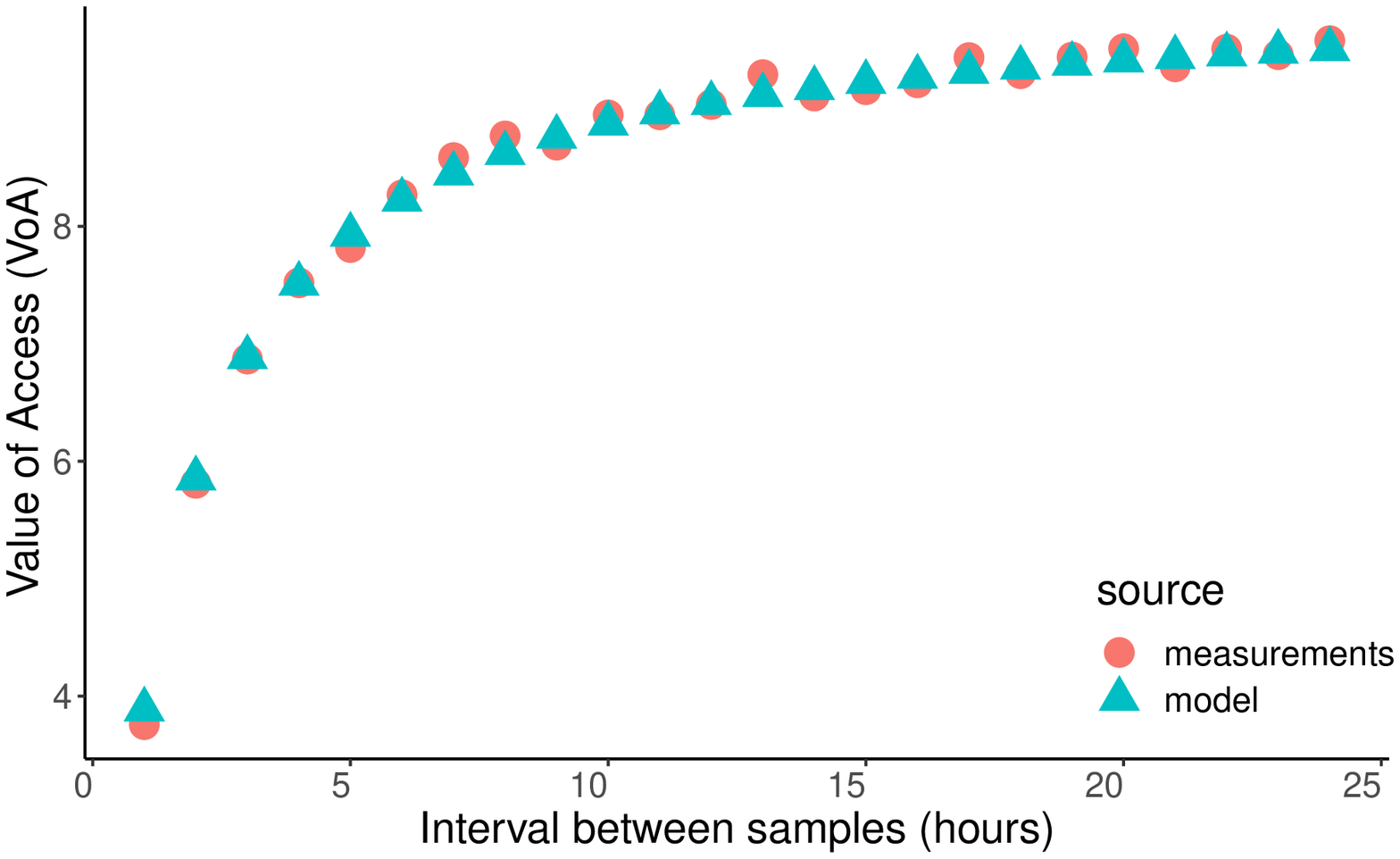}  \\
     (a) & (b) 
     
\end{tabular}
    \caption{(a) Total number of publications per day of selected pages and (b) parameterization and validation of the model with Facebook API data  }
    \label{fig:totalpubs}
\end{figure*}

The proposed analytical model was validated in a two-step process using Facebook data. First, we assess the accuracy of the model in a trace-driven simulation using only the posts created by a set of selected sources. Then, we incorporate the effect of the Facebook News Feed algorithm~\cite{News_Feed} by measuring the VoA directly at the browser's of a set of virtual users.

During the experiment, we selected the publications of the pages of the candidates for the Presidency of the Republic of Brazil on Facebook, as well as those of their respective political parties, during the first round of the electoral campaign of the presidential elections of 2018. We also selected a set of $30$ representative Brazilian's media outlets.   The data were obtained both through the Facebook API and through  a browser extension named Facebook Tracking Exposed
~\cite{fbtrex}. We collected  the publications created between September 27, 2018 and October 10, 2018. The  pages considered, and their respective addresses, are summarized in the Appendix~\ref{sec:pages}.

\subsection{Trace-driven FIFO simulations without Facebook News Feed filtering}
\label{sec:validation}

We start by validating the model before the influence of the News Feed algorithm. To this aim we envisioned collecting all the publications from the selected sources during the experiment period, through the Facebook API. Unfortunately, we could not obtain posts created by all sources because the Facebook API only retrieved  posts from the candidates and their respective political parties during this period. For this reason, we restrict to only those set of publishers in the trace-driven experiment.

Figure~\ref{fig:totalpubs}(a) shows the total number of publications per day for all selected pages during the experiment period.
The aggregate  number of publications by all sources varied between 30 and 193 publications per day. 
The number of publications ranged from a minimum of 4 to a maximum of 222 posts per page.

After selecting the sources, the behavior of a FIFO timeline of size $ K = 10 $ was simulated using the collection of the considered posts. Each snapshot is an instance of a  timeline  of size $ K$. The simulation consists of building a sequence of snapshots.  A snapshot is built  selecting a post  uniformly at random from the available posts and the  $ K-1 = 9 $ immediately preceding posts. We refer to the selected post as the \emph{reference post} of the snapshot.  For further details on the behavior of FIFO timelines, we refer the reader to~\cite{PEVA2018}.

Given the desired sample interval, $ 1 / \mu $, measured in hours, we determined the number of snapshots we took from the system. If $\mu=1$, for instance, in a period of 14 days we draw $ 24 \times 14  $ snapshots. The snapshots are then ordered in time, based on their reference posts, and from this list of snapshots we assess the value of each access (VoA).

In the experiments considered, $ 1 / \mu $ ranged from $ 1$ to $24 $. For each value of $\mu$, $30$ simulation rounds were performed.
The results are shown in Figure~\ref{fig:totalpubs}(b). The red circles are the data predicted by the model through equation~\eqref{eq:NewImpressions}, with $ \lambda = 4.487 $ and $ K = 10$. The blue triangles are the averages of the values obtained in the simulations. The deviation from the simulation is due to the fact that posts are created following a real (possibly non-Poisson) generation process. Nonetheless, this small difference between the data predicted by the model and the data obtained through simulation, indicates the Poisson process is a good approximation for the generating process.  Note that as $ 1 / \mu $ increases, VoA tends to the asymptotic value given by  $ K = 10 $, that is, the timeline will be filled completely with new information, with high probability, and the marginal gains due to an increase in the access rate are smaller.

\subsection{Facebook Tracking Exposed (FTE) experimental setup}

Next, our goal is to assess the effect of the Facebook News Feed personalization algorithm on the VoA. For this purpose, we ran an experiment  similar to the one presented in~\cite{PEVA2018}. We created $5$ virtual users, that will be referred to as bots. The bots followed \emph{all} the thirty selected \emph{pages}, most of them being media pages (see Appendix~\ref{sec:pages}), in addition to  the political candidates and their parties.   

Each  bot   kept  open an Internet   browser window  (Firefox or Chrome)  accessing the Facebook page.  The bots were prepared to collect data on the posts to which they were exposed using the Facebook Tracking Exposed~\cite{fbtrex}  browser extension.
The extension auto-scrolls the Facebook page for three minutes at scheduled moments.  Every auto-scroll produces a  set of impressions which are stored at a local database.  Each collection of posts is referred to as a \emph{snapshot}. 
\begin{enumerate}
    \item One bot was scheduled to collect snapshots every ten minutes. We refer to this user as a \emph{high} sampling user.  
    \item The \emph{regular} bot collects posts every hour. There are four regular bots.
\end{enumerate}
Each post appearing  in a snapshot counts as  a post  \emph{impression}.  
At each bot, 
Facebook Tracking Exposed collects all impressions and stores their corresponding publisher,  publication  time, impression time and impression order.

\label{sec:ftesetup}

\subsection{Experimental results accounting for Facebook News Feed  filtering}
\label{sec:NF_measurements}

Next, we report the results obtained using the methodology described in Section~\ref{sec:ftesetup}. 

\begin{figure*}
     \includegraphics[width=0.95\textwidth]{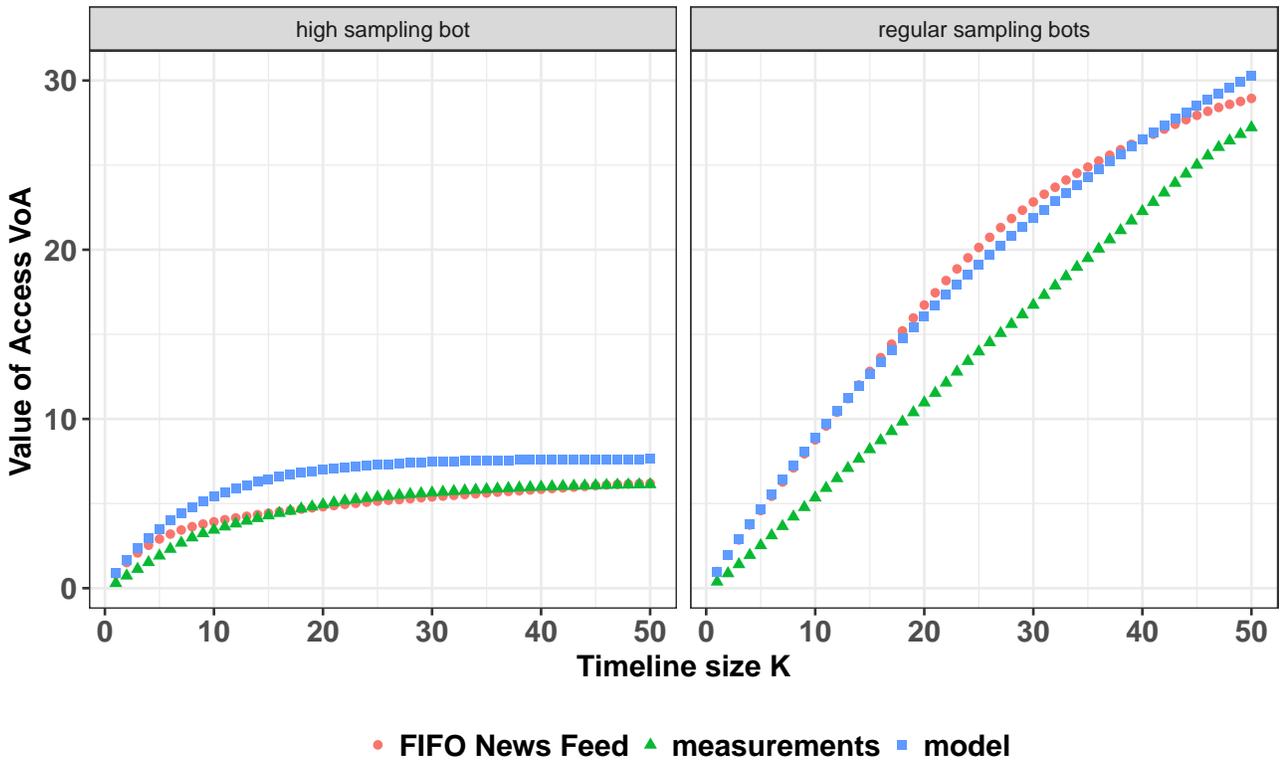}
    \caption{Model (squares) compared against the measured VoA  (triangles) and against a News Feed following a chronological order (circles).  }
    \label{fig:Fbexperiments}
\end{figure*}

Figure~\ref{fig:Fbexperiments}(a) shows the average VoA as a function of the timeline size for the two types of bots. To generate that curve from the collected measurements, for each point $K$ in the abscissa, we truncated the snapshot size to $K$ impressions, i.e.,  all the impressions that appeared in positions greater than $K$ were not considered in the VoA calculation.

We also compared the model against an alternative approach to Facebook News Feed: a timeline following a reverse chronological order, i.e., a FIFO News Feed.  To this aim, for each snapshot taken, we changed the order of the impressions such that most recent posts belonging to that snapshot appear at the topmost positions of the timeline's snapshot.

In each subfigure of Figure~\ref{fig:Fbexperiments}, the VoA predicted by the model for each value  $K$ in the abscissa is compared against the measured VoA and against the FIFO timeline for the two types of bots. At the high sampling bot, both model and measurements have similar shapes, with model predicting higher VoA than observed in measurements. For small values of $K$, i.e. $K < 5$, the FIFO News Feed and model are very close, while for larger values of $K$, i.e. $K>10$, the FIFO News Feed approaches the News Feed measurements. 

At the regular sampling bots 
the model again typically predicts a VoA higher than the measurements of Facebook News Feed. After reordering the posts, the VoA of the FIFO News Feed is accurately captured by the model. 
Therefore, the difference in the VoA as predicted by the  proposed model and the measurements is attributed  to News Feed reordering. This finding reveals that Facebook Facebook creates a \emph{position bias}~\cite{Bias_on_the_web} by showing repeated posts and is consistent with the conclusions of~\cite{PEVA2018}.
Note, for instance, that when $K=10$ it is expected according to the proposed model that the timeline can be fulfilled with new posts, i.e. VoA$=10$, but the Facebook News Feed recommends only $5$ new posts on average. To reach a  VoA of $10$, a Facebook user must scroll $20$ posts on average.  

Due to reordering, Facebook News Feed algorithm shows repeated posts, in detriment of newer ones, at the top of the timeline. One possible reason for this behavior is that Facebook shows the same post multiple times to assure that users will more likely see a post.  To overcome this effect, users must scroll their timelines beyond the repeated posts to see new contents and obtain all the value of the access.  Another possible reason is that Facebook strategically reduces the VoA as a way to increase the users' access rate.  It worth noting that the position bias has important consequences on the behaviour of users as there is a strong correlation between post positions and click rates~\cite{Bakshy2015,message}.

When $ 15< K < 35$, the model underestimates the FIFO News Feed VoA. According to equation~\eqref{eq:cond_blocking},  if $A>K$ only the $K$ most recent posts enter the timeline, while the older ones are dropped by the social media.   In reality, however, those posts can appear in future accesses.  This effect becomes more noticeable in the region wherein VoA  increases  non-linearly with respect to $K$. In the linear region, i.e. $VoA=K$, it is expected that all snapshots are filled with $K$ new posts on average. However, beyond this region  posts that are dropped in an access may be necessary to fill  timelines in future accesses, if  $A<K$ in those latter accesses. In future work, we plan to improve the model by incorporating a buffer to handle bursts of posts through an $M/G/K$ with bulk services. 

\subsection{Post overlap}

Next, we evaluate the amount of overlap between the posts viewed by the bots. This analysis objectives are two-fold: first the overlapping serves as an hypothesis check to confirm that by accessing the social network less often the bots indeed collected roughly
a subset  of the posts observed by the  bots with larger access rates. Second, the overlapping can serve to verify if Facebook News Feed algorithm has influence over VoA.

Table~\ref{tab:overlap_reg} shows the overlap between the impressions observed by the high sampling bot 
and all four regular bots. The notation goes as follows:
\begin{itemize}
    \item a post in $XY$ was viewed by the two bots,
    \item a post in $X\bar{Y}$ was viewed only by X,
    \item a post in $\bar{X}Y$ was viewed only by Y,
    \item a post in  $\bar{X}\bar{Y}$ was not seen by any of the two bots.
    \end{itemize}

The high sampling bots viewed at least $80\%$ of the regular sampling bots' impressions, indicating that regular bots impressions are a subset of the impressions of the high sampling bot.  

\begin{figure}
     \includegraphics[width=0.5\textwidth]{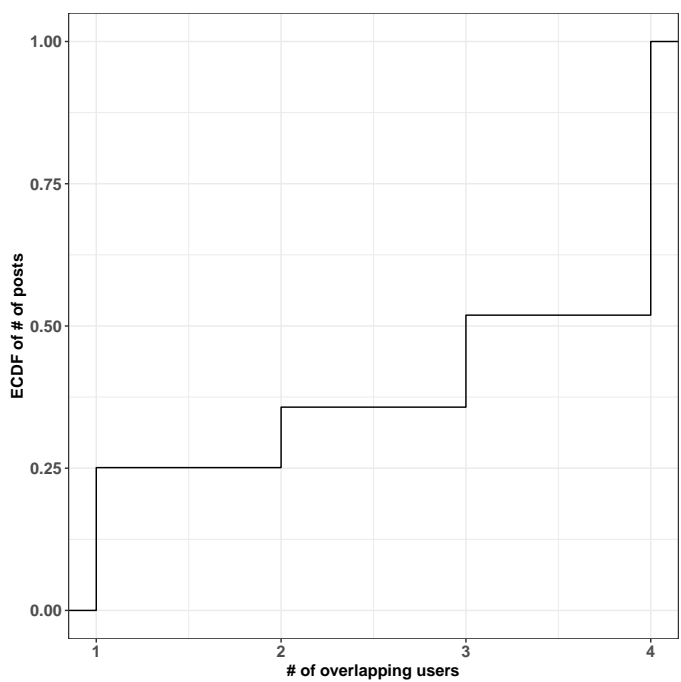}
    \caption{ECDF for the number of posts viewed versus the number of overlapping users   }
    \label{fig:Overlapping}
\end{figure}

Table~\ref{tab:overlap_reg_reg} shows that the overlap between the four regular bots ranges between $74\%$ to $89\%$. Figure~\ref{fig:Overlapping} shows the ECDF of the number of posts that were viewed by $x$ overlapping bots. It can be seen that almost half of the posts were viewed by all the 4 bots and one quarter were viewed by only on bot. It is worth noting that all bots follow the same set of pages and are scheduled to access Facebook simultaneously, thus, these overlapping patterns help to 
detect the impact of the Facebook News Feed algorithm in the bots impressions. These results together with the findings reported in Section~\ref{sec:NF_measurements}  indicate that Facebook News Feed personalization algorithm reduces the VoA when deciding which posts will be displayed to the users, particularly in the topmost positions.


 Tables~\ref{tab:overlap_reg} and~\ref{tab:overlap_reg_reg}  also indicate  that the low sampling bots  collected posts that were not seen by any other bot.  This  indicates that studying the impact of the rate at which the network is accessed is non-trivial when we account for the filtering effect of the News Feed.  A number of factors such as the time of the day, the sex of the bot and the dynamic evolution of number of likes to a given post are some of the various variables that affect the popularity of posts, which in turn impact the VoA.  Although Tables~\ref{tab:overlap_reg} and~\ref{tab:overlap_reg_reg} indicate that an analysis which neglects such second order effects is already valuable, they also point to the fact that there is significant space for improvement in the analysis of the VoA beyond the simple models considered in this paper.

\begin{table}[]
\centering
    \caption{Post overlap between the high sampling bot and the regular sampling ones}
    \begin{tabular}{cc}
\begin{tabular}{l|l|l}
                         & $R_1$    &  $\bar{R_1}$ \\
\hline
$H$                        & 8205 & 2778                     \\
$\bar{H}$ & 651 & 2679                    
\end{tabular}
&\begin{tabular}{l|l|l}
                         & $R_2$    &  $\bar{R_2}$ \\
\hline
$H$                        & 7242 & 3751                     \\
$\bar{H}$ & 1767 & 1563                    
\end{tabular}
\\
\\
(a) Regular I & (b) Regular II \\
\\
\begin{tabular}{l|l|l}
                         & $R_3$    &  $\bar{R_3}$ \\
\hline
$H$                        & 8381 & 2602                     \\
$\bar{H}$ & 1128 & 2202                    
\end{tabular}
&\begin{tabular}{l|l|l}
                         & $R_4$    &  $\bar{R_4}$ \\
\hline
$H$                        & 8111 & 2872                     \\
$\bar{H}$ & 801 & 2529                    
\end{tabular}
\\ \\
(c) Regular III & (d) Regular IV
\end{tabular}
\label{tab:overlap_reg}
\end{table}

\begin{table}[]
\centering
    \caption{Post overlap between the regular sampling bots}
    \begin{tabular}{cc}
\begin{tabular}{l|l|l}
                         & $R_2$    &  $\bar{R_2}$ \\
\hline
$R_1$                        & 6776 & 2080                     \\
$\bar{R_1}$ & 2223 & 3224                    
\end{tabular}
&\begin{tabular}{l|l|l}
                         & $R_3$    &  $\bar{R_3}$ \\
\hline
$R_1$                        & 6776  & 2080                     \\
$\bar{R_1}$ & 1453 & 4004                    
\end{tabular}
\\
\\
(a) Regular 1,2 & (b) Regular 1,3 \\
\\
\begin{tabular}{l|l|l}
                         & $R_4$    &  $\bar{R_4}$ \\
\hline
$R_1$                        & 7850  & 1006                     \\
$\bar{R_1}$ & 1062 & 4395                   
\end{tabular}
&\begin{tabular}{l|l|l}
                         & $R_3$    &  $\bar{R_3}$ \\
\hline
$R_2$                        & 6855 & 2144                     \\
$\bar{R_2}$ & 2654 & 2660                    
\end{tabular}
\\ \\
(c) Regular 1,4 & (d) Regular 2, 3\\
\\
\begin{tabular}{l|l|l}
                         & $R_4$    &  $\bar{R_4}$ \\
\hline
$R_2$                        & 6641 & 2358                     \\
$\bar{R_2}$ & 2271 & 3043                   
\end{tabular}
&\begin{tabular}{l|l|l}
                         & $R_4$    &  $\bar{R_4}$ \\
\hline
$R_3$                        & 7734 & 1775                     \\
$\bar{R_3}$ & 1178 & 3623                   
\end{tabular}
\\ \\
(e) Regular 2,4 & (f) Regular 3, 4\\
\end{tabular}
\label{tab:overlap_reg_reg}
\end{table}

\section{Finding the optimal sampling rate}
\label{sec:control}

In this section we pose the optimal access rate problem. We consider users who have a cost to access their social networks. This cost reflects, for instance,  the users attention budget, as well as the  energy spent from the batteries of   mobile devices and possible monetary costs associated to  using data plans for cellular access.
Let $ c $ be the cost incurred by a user per access, measured in the same units as the VoA.


The \emph{utility} per  time unit  experienced by a user is given by the expected number of new impressions per view ($ \overline {V} $), times the rate of views per  time unit ($ \mu $), minus  the cost $ c \mu $ required to recover these impressions. Therefore,
\begin{equation}
    U(\overline{V},\mu)=\mu \overline{V}-c\mu.
    \label{eq:utility_N}
\end{equation}

Substituting \eqref{eq:NewImpressions} into  \eqref{eq:utility_N}, we express the utility as a function of  $ \lambda $ and $ K$:

\begin{equation}
    U(\mu)= \lambda \bigg(1- \bigg(\frac{\lambda}{\lambda+\mu}\bigg)^K\bigg)-c\mu.
    \label{eq:utility}
\end{equation}

Therefore, given $ \lambda $ and $ K $, the problem of determining the optimal access rate to a  social network consists of  maximizing the utility by adjusting the access rate $  \mu $,
\begin{equation}
    \max \quad U(\mu), \quad \mu \geq 0, 
    \label{eq:problem}
\end{equation}
where $ U (\mu) $ is given by  \eqref{eq:utility}.
The utility is maximized when its partial derivative with respect to the control variable is equal to zero, i.e., when $ {\partial U} / {\partial \mu} = 0 $.
The partial derivative $ {\partial U} / {\partial \mu} $ is given by

\begin{align}
\frac{\partial U}{\partial \mu}=K \bigg(\frac{\lambda}{\lambda+\mu}\bigg)^{K+1}-c.
\end{align}

Then, letting  $ {\partial U} / {\partial \mu} = 0 $ we obtain the optimal access rate,~$\mu^{\star} $,

\begin{align}
\mu^{\star} &= \lambda \bigg( \frac{\sqrt[K+1]{K}}{\sqrt[K+1]{c}}-1 \bigg).
\label{eq:optimal}
\end{align}

Next, we numerically evaluate  the behavior of the optimal access rate, as a function  of the different system parameters.


\section{Numerical Examples}
\label{sec:numerical}

 \begin{figure*}
    \centering
\begin{tabular}{cc}
     \includegraphics[width=0.5\textwidth]{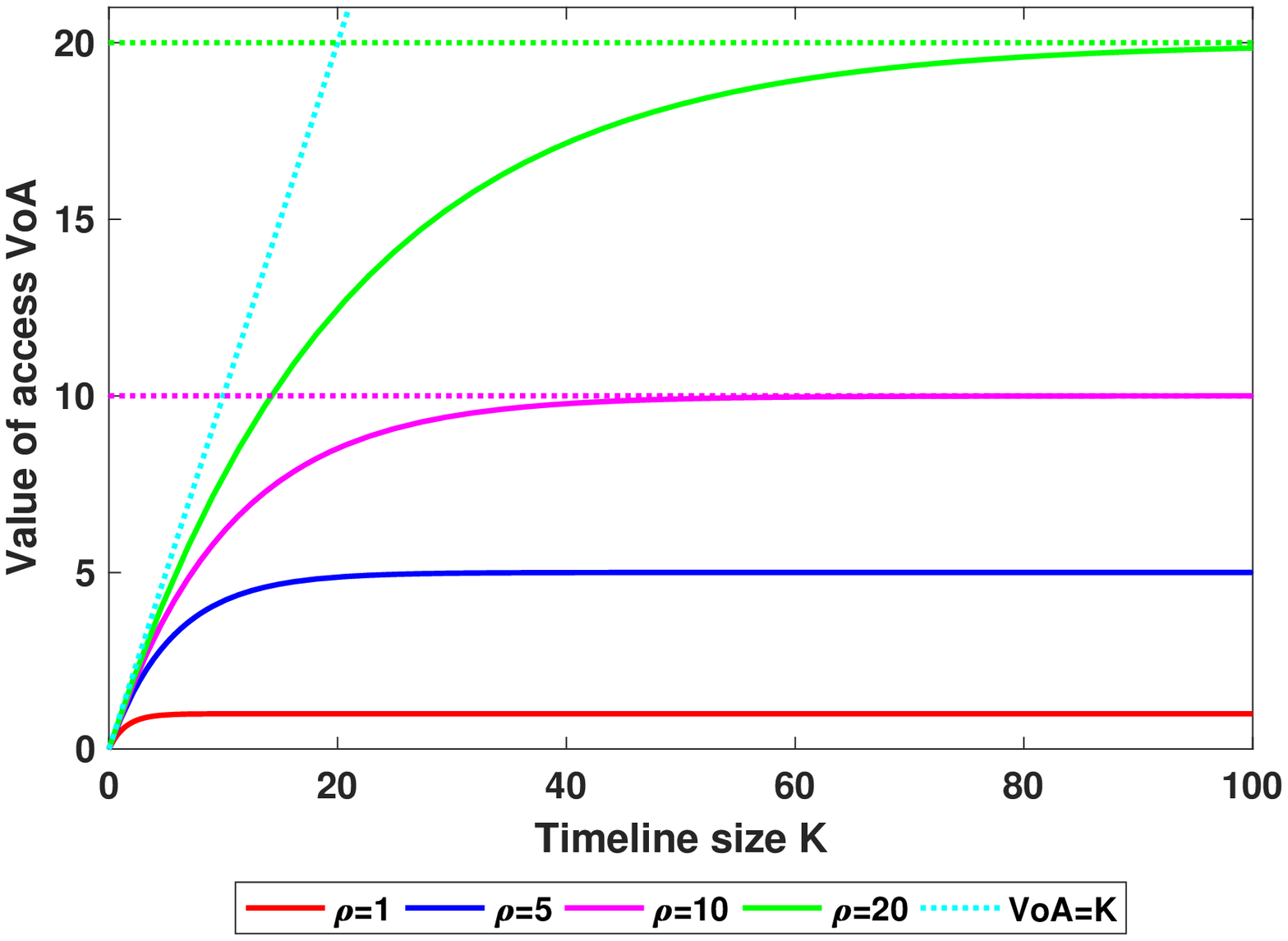} &      \includegraphics[width=0.5\textwidth]{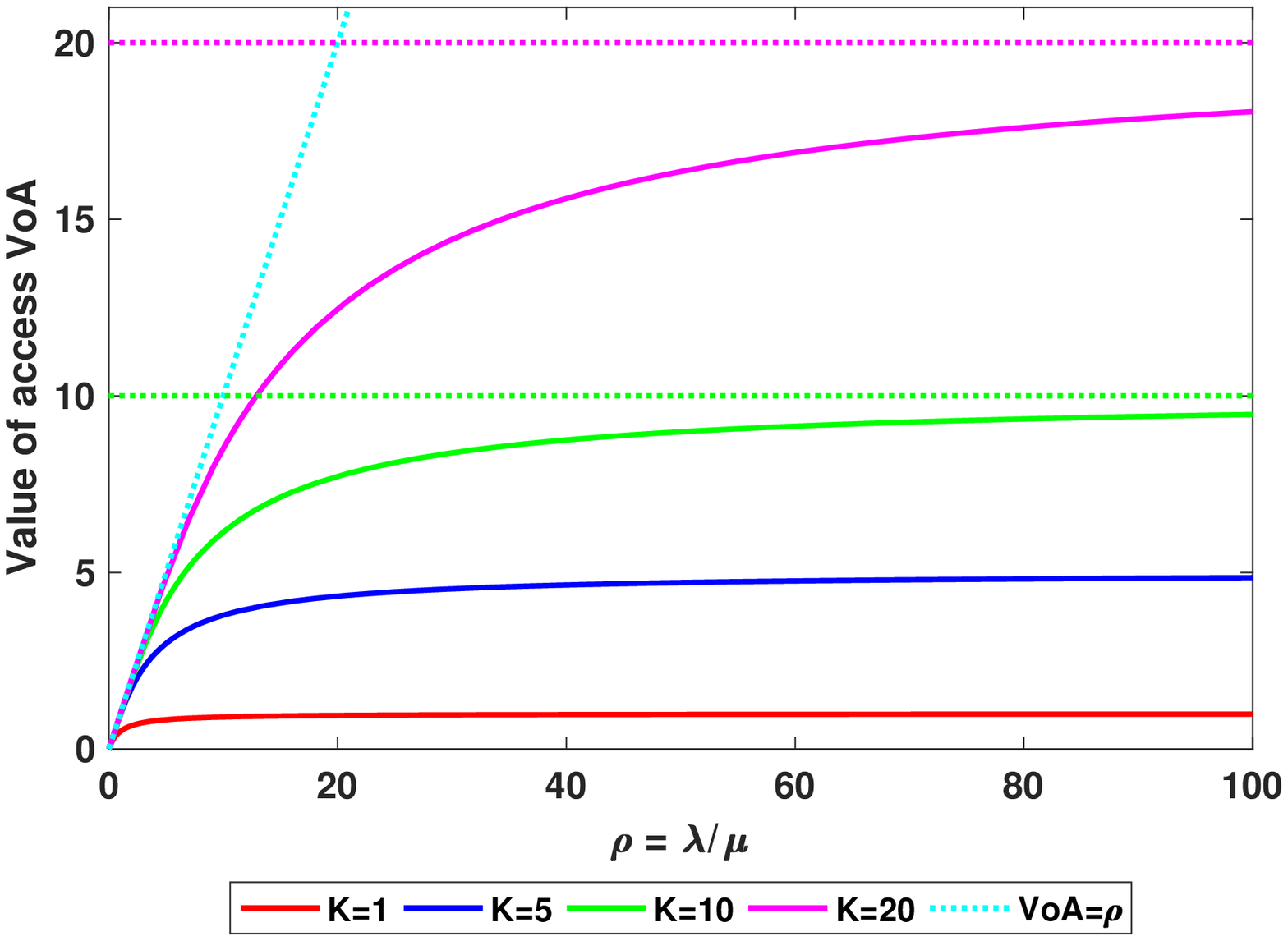}  \\
     (a) & (b) 
     
\end{tabular}
    \caption{Sensitivity analysis: impact of different parameters ($ K, \rho $) on the value of an access (VoA).} 
    \label{fig:VoA}
\end{figure*}

In this section we present numerical examples obtained from the proposed model. Our objectives are $ (i) $ to perform sensitivity analysis of the metrics of interest given the parameters of the model (proposed in Section \ref{sec:model}) and $ (ii) $ to illustrate the sensitivity of the solutions of the optimal access rate problem  (introduced in Section \ref{sec:control}) to its parameters.

\subsection{Impact of timelime size $K$ and ratio $\rho = \lambda/\mu$}

Next, we evaluate the marginal increment of VoA resulting from  users  scrolling down  their screen to capture more access impressions, i.e., we evaluate the impact of an increase in  $ K $ on VoA.
Figure~\ref{fig:VoA}(a) shows the impact of $ K $ on VoA, for a fixed value of $ \rho $. At the limit when $ K \to \infty $, $ \overline{V} (\rho, K) = \rho $.
Note also that in all the scenarios considered the convergence for values close to the asymptotic value occurs for $ K \leq 100$.

For example, when $\rho = 10 $, that is, when the access rate is 10 times smaller than the publication rate, $\overline{V} \approx \rho$ for $K \ge 40$. Thus, even if a user accesses the network infrequently, at $ \mu = \lambda / 10 $, he can still absorb almost every novelty provided he can recover and read $ K = 40 $ posts per access.

When users capture $ K $ impressions per access, the maximum reachable VoA is $ K $.
How does the VoA actually experienced by users compare to the maximum achievable VoA?

In the  dotted line shown in Figure~\ref{fig:VoA}(a) we have the VoA corresponding to the situation in which all received publications are new. Figure~\ref{fig:VoA}(a) shows that even for small values of $ K$, e.g., $K = 10$, the experienced VoA already significantly distances itself from $K$. This effect is caused both by fluctuations in the number of publications between accesses and by variations in the interval between accesses. \emph{Note that if the accesses and the publications are deterministic and synchronized, the dotted and full curves will overlap.
 However, publications arise organically, due to exogenous factors, and dealing with unpredictability is an intrinsic aspect of the problem}.

Figure~\ref{fig:VoA}(b) shows the increase of VoA as a function of $ \rho$.
When $ \rho \to \infty $, $ \overline{V} (K,\rho) = K$.
Note that the greater the value of $ K$, the greater the value of $ \rho $ required to reach the asymptotic value $ \overline{V} = K $. In fact, the greater
the value of $ K $, the more active the sources need to be in order to fill in the timelines of users with novel impressions.

\begin{figure*}[t]
\footnotesize
    \centering
\begin{tabular}{cc}
     \includegraphics[width=0.5\textwidth]{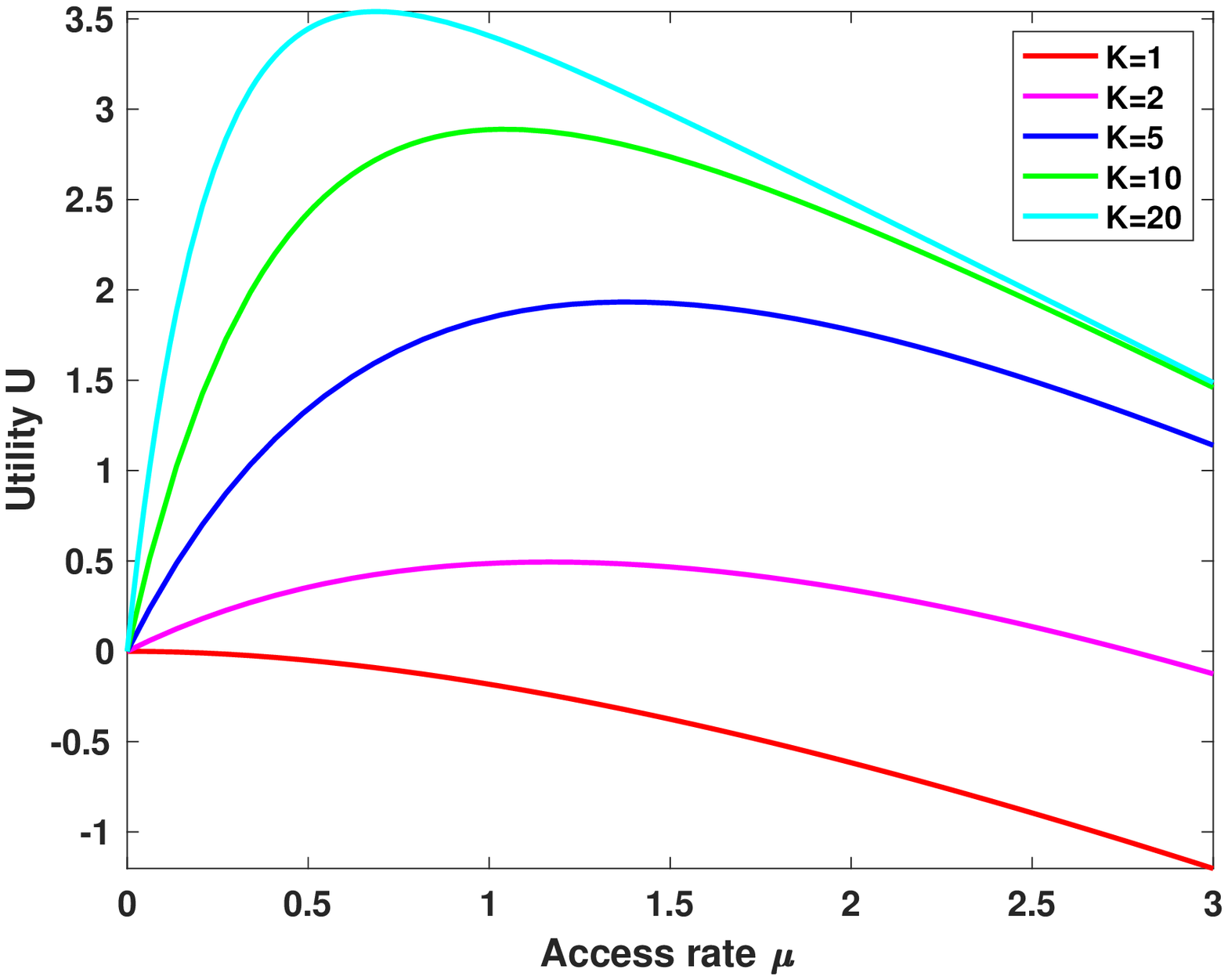} &      \includegraphics[width=0.5\textwidth]{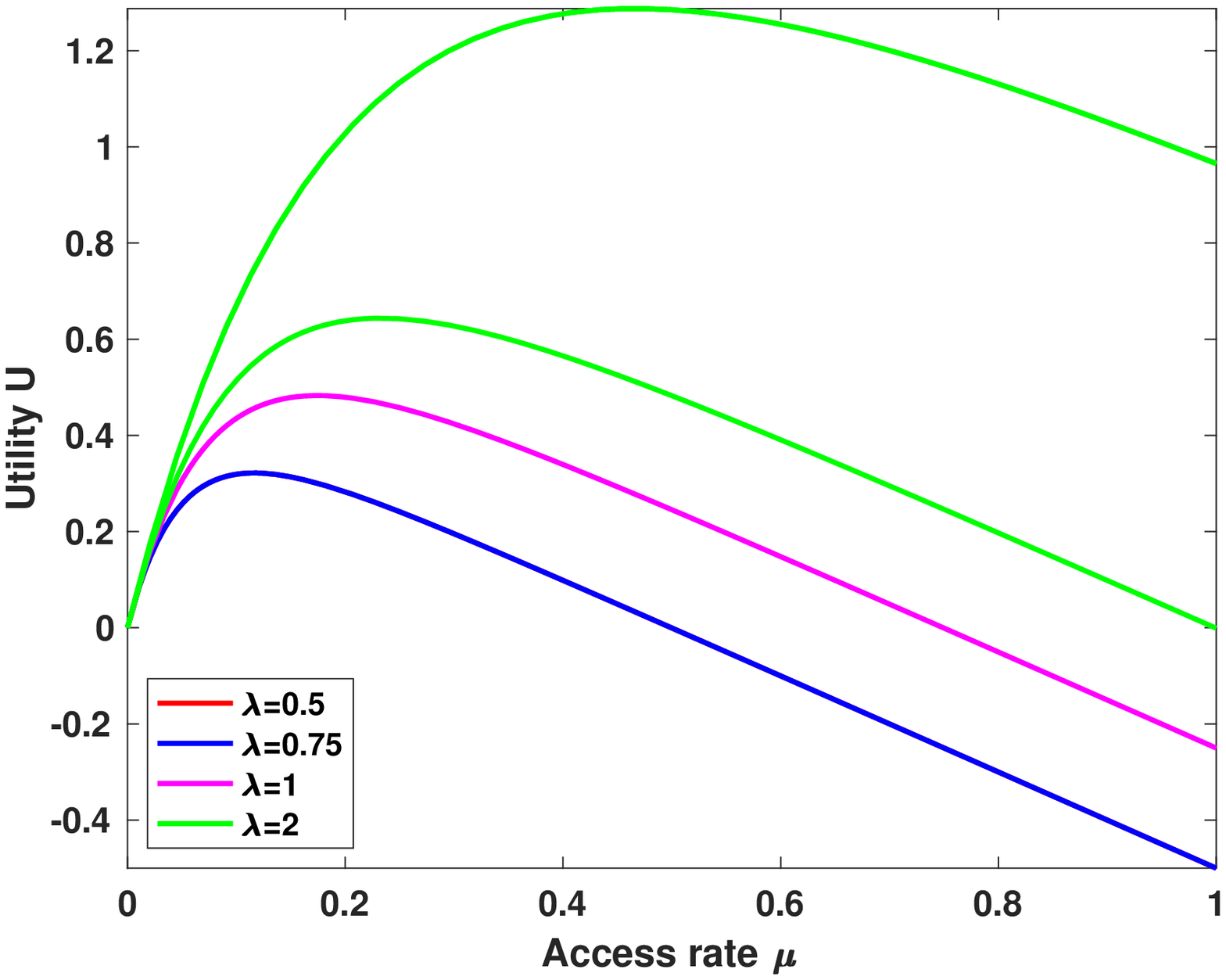}  \\
     (a) & (b) \\
     \includegraphics[width=0.5\textwidth]{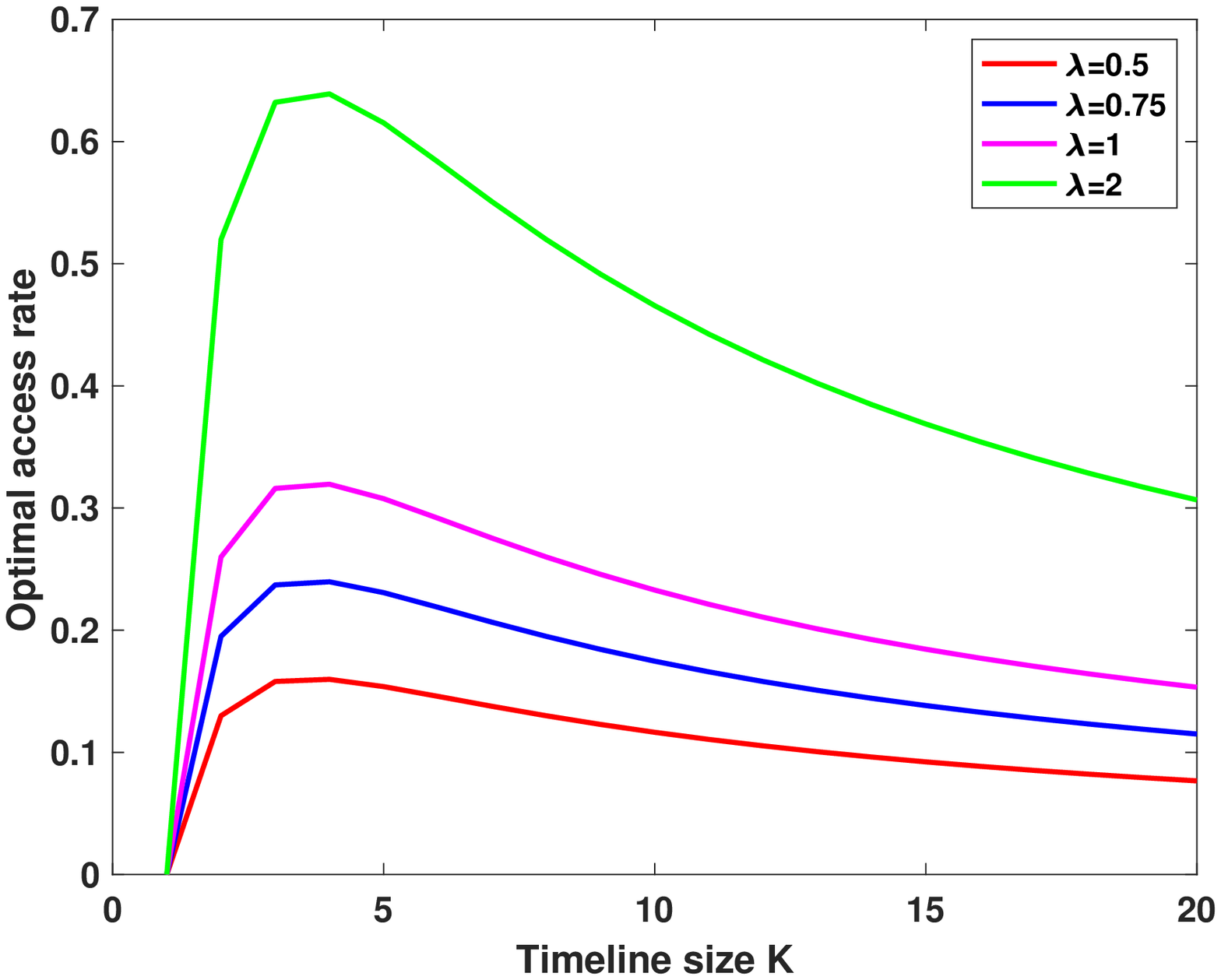}     &      \includegraphics[width=0.5\textwidth]{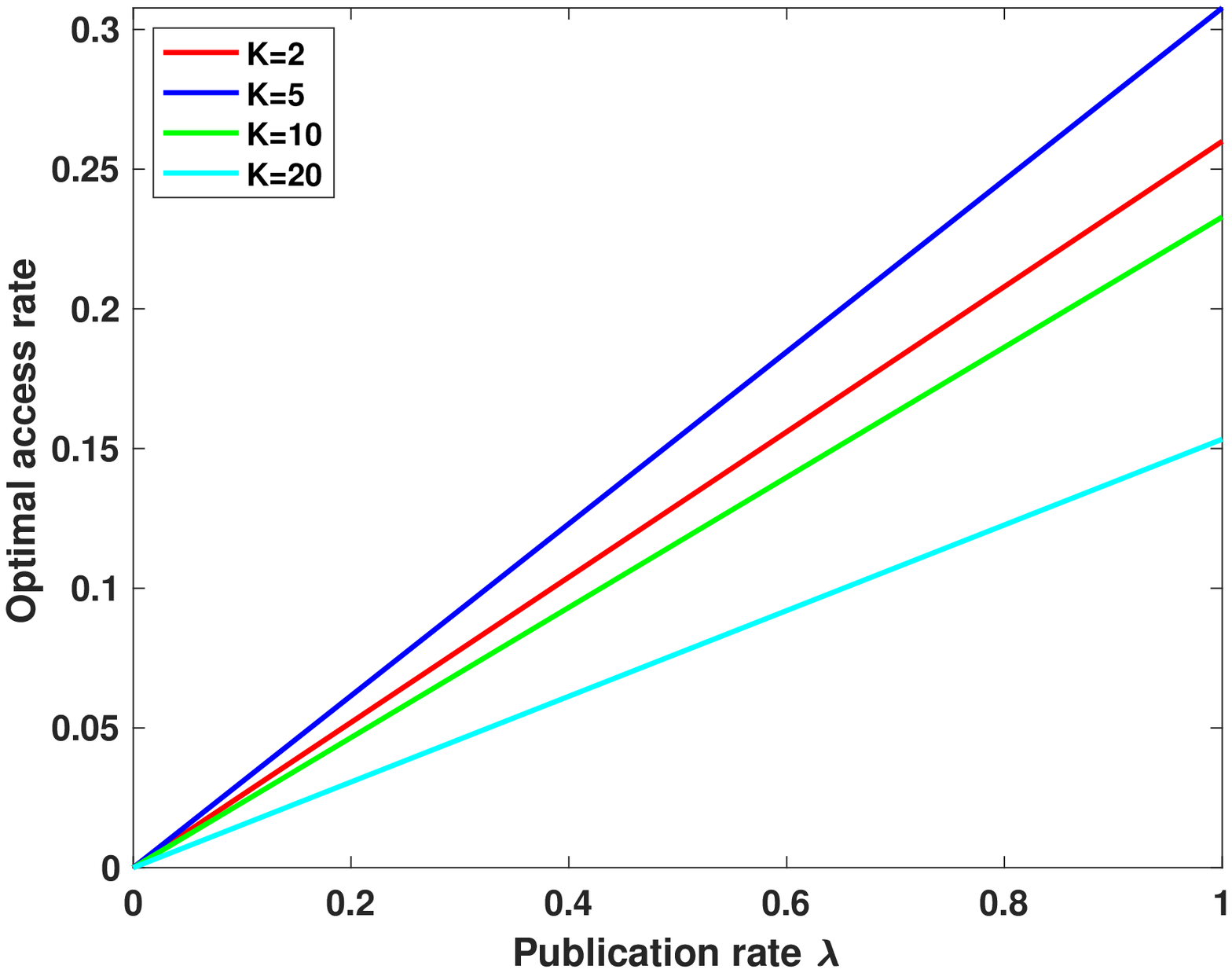} \\
     (c) & (d)
\end{tabular}
    \caption{Sensitivity analysis: impact of different parameters ($ K, \lambda,  \mu $) on users' utility ((a) and (b)) and optimal access rate ((c) and (d)).}
    \label{fig:Utilidade}
\end{figure*}

\subsection{Optimization problem}

Figure~\ref{fig:Utilidade}(a) shows the utility experienced by a user per time unit  as a function of the access rate $ \mu $, for five values of $ K $. In this figure, $ \lambda $ has been parameterized with the data returned by the Facebook API as described in Section ~\ref{sec:validation}. The cost $ c $ is fixed equal to $ 1 $. This figure shows that, given the other  parameters constant, the higher the value of $ K $, the lower the optimal access rate. For example, when $K$ increases from 2 to 20, $\mu^{\star}$ decreases from $ 1.1663$ to $0.688$ accesses/hour. For a given pair $ (\lambda, \mu)$, the greater the value of $ K$, the greater the utility. This is because we assume in our model that the cost of an access is independent of $ K $. More complex cost functions will be  subject of future work.

Figure \ref{fig:Utilidade}(b) shows the utility experienced by a user per time
 unit as a function of the access rate $ \mu $ for four values of $ \lambda $ (with $ K = 10$). In this case, as  $ \lambda $ grows the optimal access rate increases. For a given pair  
 $ (K,  \mu)$, the utility increases with respect to  $ \lambda $. In fact, more active sources will favor an increase of VoA, remembering that such value is limited by~$K$.

Figure \ref{fig:Utilidade}(c) shows the optimal access rate $ \mu^{\star} $ as a function of $K$. Note that for values of $ K $ close to one,  $K \approx 1$, as $ K $ increases, $  \mu^{\star} $ increases  as the timeline can still be filled  with additional novel posts. However, if $ K $ is further increased the system eventually transitions to a regime wherein  as $K$ grows the optimal sampling rate $ \mu^{\star}$ decreases. This is because  the probability of observing repeated impressions eventually becomes non-negligible.

Figure~\ref{fig:Utilidade}(d) shows the optimal access rate $ \mu^{\star} $ as a function of~$\lambda$, for $K \in \{2,5,10,20\}$. This figure illustrates the linear behavior of $ \mu^{\star} $ as a function of $ \lambda $ (see~\eqref{eq:optimal}). Next, consider Figure~\ref{fig:Utilidade}(d)  for a  given value of $ \lambda$.  Note that when $K$ grows from $K=2$ to $K=5$, $ \mu^{\star} $ increases as  $ K $ grows,  as the timelines are still not saturated by  the new posts being created. As $K$ increases  in the range from $K=5$ to $K=20$, however, $\mu^{\star} $  decreases. This is because in this regime the timelines are already saturated, and the probability to receive repeated impressions increases. In this case, \emph{if users can  consume more publications per access, they can access the network less frequently (see also Fig.~\ref{fig:Utilidade}(c))}.

\subsection{Practical implications on social network design}
The numerical examples presented in this section illustrate that VoA can be 
limited  either by $ \rho $ or $ K $. The system  then goes through two operating regimes:
\begin{itemize}
    \item Consider the regime in which $ \overline{V} \approx \rho $. We consider a  fixed and given value of $ K$, and note  that in this regime the increase of $ K $ typically does not result in an increase of $ \overline{V}$. If the user is able to consume $ K $ new impressions per access, he does not need to use social media filters;
    \item  In the regime wherein $ \overline{V} \approx K $ the user is potentially losing many new publications. In that case, he needs the help of the social network's recommendation system to filter out the most relevant publications.
\end{itemize}
Understanding these two regimes helps in the design of  online social network timelines.  Characterizing the regime in which recommendation algorithms such as the Facebook News Feed are needed to add value to users is key to determine when and if filters play a more prominent role and to assess their impact on the fairness and accountability of online social networks.
\section{Related work}
\label{sec:related}

The rate at which users access their online social networks has implications that range from Psychology and Economics to network infrastructure and performance.   Next, we discuss prior work on each of those three domains. 
Then, we also  briefly  survey previous efforts on  analytical models for online social network timelines.
\emph{We are unaware of previous work considering analytical models to estimate the optimal rate at which users should access their online  social networks, and that has parameterized such models with real data. }

\subsection{Psychology and addiction} \label {sec: psi}

The appropriate rate for users to access online social networks is a theme that has been widely discussed in the sphere of Psychology ~\cite{andreassen2012development, Hussain2019,salehan2013social,HONG2014597,ryan2014uses}.
According to Sean Parker, founder of Napster and former president of Facebook, \emph{Facebook is a system designed to exploit human vulnerability}~\cite{SeanParker}. 
Accordingly,  
Tristan Harris, a former Google employee, claimed that \emph{technological artifacts can hijack people's minds}~\cite {TristanHarris}.
 Although in this work our focus is exclusively related to technical questions regarding the amount of novelty users get each time they access a social network, the psychological implications of this work are evident. In particular, if
 users know in advance that they will receive little news when accessing the network at any given moment, they will have more incentives to further space their access. 
 
 \subsection{Attention economy and ads} \label{sec:econ}
The rate at which users access online social networks has immediate economic
implications, given that advertisements generate profits based on \emph{click rate}. The click rate, in turn, is strongly influenced by the rate at which users access their online social networks. 
If advertisers have access to information on the recommended rate for users to access 
their online social network throughout the day, they can better decide when to pay for certain advertisements to be displayed, 
e.g., depending on the target audience ~\cite{rendle2012social, jiang2013optimally}. 

The dynamics of collective attention and the way that popularity of topics evolves over time was studied in~\cite{Lorenz-Spreen2019}.  Combining these dynamics with user's behaviour expressed by the sampling rate
and  filtering algorithms~\cite{PEVA2018,Milan2015}, one
can derive insights on the dynamics of  topic popularity and production of information~\cite{Ciampaglia2015}, particularly  when there is contention for visibility and attention.

\subsection{Infrastructure and age of information (AoI)}
\label{sec: aoi}
The rate at which users access online social networks has a direct impact on the traffic passing through the network. Since social networks are currently one of the largest sources of traffic, recommendations on the rate at which users should access their 
pages can have cascading effects on the network as a whole. The age of information (AoI) has gained considerable visibility in the scientific community, and we envision that the contributions of this work are relevant in this new domain~\cite{kosta2017age,Chaintreau:2009,ForeverYoung, yates2019age,wu2017optimal,Kadota:2019,Bedewy:2019,Ayan2019}.

\subsection{Modeling and evaluation of timelines' performance} \label{sec:timelines}
Facebook's News Feed was modeled as a FIFO queue in~\cite{eduardoasonam}. In~\cite{PEVA2018} it has been shown that FIFO timelines are a particular case of  time-to-live (TTL) timelines~\cite{dehghan2016utility}.  
TTL timelines  are versatile and flexible,
 allowing the reordering of publications. 
 In this paper, 
 we build on top of~\cite{PEVA2018} and~\cite{giovanidis2019performance} to  account for the impact of  the user's access rate, opening up new perspectives of analysis and enabling the formulation of the problem of optimal access rate of users.

\section{Discussion}
\label{sec:discussion}

The ideas presented in this work can be implemented in an intermediate layer between the user and the social network, e.g., through an application that consumes the Facebook News Feed and learns the parameters of the model proposed in this paper. Platforms like Gobo~\cite{gobo} already act as a \emph{middleware} between Facebook and users, and can serve as a basis for implementing the ideas presented here.

The results contained in this paper do not consider all the impact that online social networking algorithms such as Facebook's News Feed can have on VoA. In future work, we intend to quantify the VoA by user profile and by social network, e.g., using platforms such Facebook and Twitter. From this characterization of the VoA, we envision to use VoA as a criterion for comparing online social networks personalization algorithms and their effects on the information diets~\cite{juhiInfoDiets}, e.g., accounting for  bots with multiple political  orientations~\cite{PEVA2018}.
\section{Conclusion}
\label{sec:conclusion}

Social networks are increasingly present in the daily lives of millions of Internet users, and a significant portion of the world's population accesses social networks multiple  times on a daily basis~\cite{WeAreSocial}. Each access to an online social network yields a certain \emph{value} to its user, which is 
a function of numerous subjective aspects.  For this reason,  measuring the \emph{value of access} (VoA) is a challenging endeavour.

In this work, we overcome the challenges
associated to the assessment of VoA by proposing a simple but quantifiable approach to estimate the VoA experienced by a user of
 an social network. In particular, we assume that this value is equal to the amount of new impressions presented by the social network to the user. Using this approach, we proposed an analytical model to quantify the VoA. The model was validate using real-world data from Facebook experiments.  
 Next, we parameterize the model using real data, and indicate how the model can be used to obtain the optimal access rate.

 The results enable us to conclude that Facebook News Feed algorithm reduces the VoA by deciding that some repeated post are more relevant to the users than other fresher and unseen posts, specially in the topmost positions. Even bots following the same set of publishers, and accessing Facebook simultaneously, experienced a personalized information diet. We believe this work is yet another step in a research agenda that aims to develop social networking platforms respecting users' preferences. This agenda necessarily involves contemplating the psychological, economic and infrastructure aspects inherent to the problem of determining the optimal rate at which users should access online social networks, which we intend to embrace in future works.

\section*{Acknowledgment}
This work was partially supported by CAPES, CNPq, FAPERJ and Petrobras, as well as by the joint Brazilian-French team THANES (jointly sponsored by FAPERJ and INRIA).

\begin{appendix}

\subsection{Terminology}

\label{sec:termino}

Next, we summarize the basic terminology used throughout this paper.
\begin{itemize}
\item\textbf{Post} is a message created by a source  
\item\textbf{Access}, also referred to as \emph{view}, is the act of refreshing a timeline
\item\textbf{Timeline} is a sorted list of posts presented to a user.  Every time a user accesses the network, a  timeline is prepared.  The timeline may contain posts that have already been presented to the user before.  Nonetheless, there are no repeated posts in a given timeline
\item\textbf{Impression} is a post appearing in a timeline of a user
\item\textbf{Timeline size} is the maximum number of elements in a timeline
\item\textbf{Access rate} is the rate at which a user accesses the network. At every access, the user can possibly receive a timeline containing novel impressions
\item\textbf{Novel impression} is an impression corresponding to a post that has not yet been presented to that user
\item\textbf{Repeated impression} is an impression corresponding to a post that has already been presented to that user
\item\textbf{Value of access (VoA)} is the number of novel impressions in a timeline 
\item\textbf{Access cost} is the cost of accessing a social network, e.g., measured in monetary units, but reflecting the attention span dedicated to the access
\item\textbf{Utility of an access} is the value of the access minus the associated costs.
\end{itemize}

\subsection{Expression for $\overline{V}$ under the exponential model}

\label{sec:simiplificaen}

From~\eqref{eq:exprndet} and \eqref{eq:NGivenTau} we have
\begin{equation}
    \overline{V}=K-\sum_{i=0}^K \frac{(K-i)}{i!}\lambda^{i}\mu
    \int_{\tau=0}^{\infty} e^{-(\lambda+\mu)\tau}\tau^{i} d\tau
\end{equation}
The solution of the integral above can be found in standard tables of integrals,
\begin{equation}
    \overline{V}=K-\sum_{i=0}^K \frac{(K-i)}{i!}\lambda^{i}\mu (\lambda+\mu)^{-(i+1)}\Gamma(i+1)
\end{equation}
Then, applying the definition of the gamma function,  $\Gamma(i+1)$,
\begin{equation}
    \overline{V}=K-\sum_{i=0}^K \frac{(K-i) }{i!}\frac{\lambda^{i}\mu}{(\lambda+\mu)^{(i+1)}}i! 
    \label{eq:withSummation}
\end{equation}
Reorganizing the terms in~\eqref{eq:withSummation},
\begin{equation}
    \overline{V}= K -\frac{\mu}{\lambda+\mu}\bigg(K\sum_{i=0}^K \frac{\lambda^i}{(\lambda+\mu)^i}-\sum_{i=0}^K \frac{i\lambda^i}{(\lambda+\mu)^i}\bigg).
    \label{eq:with2Summation}
\end{equation}

After algebraic manipulation, and noting that the two summations in \eqref{eq:with2Summation} have closed expressions,  we obtain a simple expression for $\overline{V}$ (see~\eqref{eq:NewImpressions}),

\begin{align}
\overline{V} = \frac{\lambda}{\mu} \left(1-\left(\frac{\lambda}{\lambda+\mu} \right)^{K} \right).
\end{align}

\subsection{List of publishers and pages}
\label{sec:pages} 

Table~\ref{tab:pages} reports the list of publishers and pages followed by the bots considered in our experimental setup.

\begin{table*}[t]
 \centering
    \caption{List of pages followed}
\begin{tabular}{ll}
Page                   & Address                                              \\
Jair Bolsonaro           & https://www.facebook.com/jairmessias.bolsonaro/       \\
PSL                      & https://www.facebook.com/PartidoSocialLiberalBR/      \\
Lula                     & https://www.facebook.com/Lula/                        \\
Fernando Haddad          & https://www.facebook.com/fernandohaddad/              \\
PT                       & https://www.facebook.com/pt.brasil/                   \\
Geraldo Alckmin          & https://www.facebook.com/geraldoalckmin/              \\
PSDB                     & https://www.facebook.com/Rede45/                      \\
Marina Silva             & https://www.facebook.com/marinasilva.oficial/         \\
Rede Sustentabilidade    & https://www.facebook.com/RedeSustentabilidade18/      \\
Ciro Gomes               & https://www.facebook.com/cirogomesoficial/?ref=br\_rs \\
PDT                      & https://www.facebook.com/pdt.org.br/                  \\
Henrique Meirelles       & https://www.facebook.com/hmeirellesoficial/           \\
MDB                      & https://www.facebook.com/MDBNacional15/               \\
Álvaro Dias              & https://www.facebook.com/ad.alvarodias/               \\
Podemos                  & https://www.facebook.com/podemos19/                   \\
Guilherme Boulos         & https://www.facebook.com/guilhermeboulos.oficial/     \\
PSOL                     & https://www.facebook.com/psol50/                      \\
Cabo Daciolo             & https://www.facebook.com/depudadocabodaciolo/         \\
Patriotas                & https://www.facebook.com/Patriota51Oficial/           \\
João Amoes               & https://www.facebook.com/JoaoAmoedoNOVO/              \\
Partido Novo             & https://www.facebook.com/NOVO30/                      \\
Quebrando o tabu         & https://www.facebook.com/quebrandootabu/              \\
Jornal O Globo           & https://www.facebook.com/jornaloglobo/                \\
BBC Brasil               & https://www.facebook.com/bbcbrasil/                   \\
BuzzFeed Brasil          & https://www.facebook.com/BuzzFeedBrasil               \\
Carta Capital            & https://www.facebook.com/CartaCapital/                \\
Mídia Ninja              & https://www.facebook.com/MidiaNINJA/                  \\
Huffington Post Brasil   & https://www.facebook.com/HuffPostBrasil/              \\
Jornalistas Livres       & https://www.facebook.com/jornalistaslivres/           \\
El Pais Brasil           & https://www.facebook.com/elpaisbrasil/                \\
Revista Forum            & https://www.facebook.com/forumrevista/                \\
Brasil de fato           & https://www.facebook.com/brasildefato/                \\
DW Brasil                & https://www.facebook.com/dw.brasil/                   \\
Observatório da Imprensa & https://www.facebook.com/Site.ObservatoriodaImprensa/ \\
Congresso em foco        & https://www.facebook.com/congressoemfoco/?ref=br\_rs  \\
Le Monde Brasil          & https://www.facebook.com/diplobrasil/?ref=br\_rs      \\
The Intercept Br         & https://www.facebook.com/TheInterceptBr/?ref=br\_rs   \\
Agência Pública          & https://www.facebook.com/agenciapublica/              \\
Spotniks                 & https://www.facebook.com/spotniks/                    \\
Poder 360                & https://www.facebook.com/Poder360                     \\
O Estado de SP           & https://www.facebook.com/estadao/?ref=br\_rs          \\
Revista Isto é           & https://www.facebook.com/revistaISTOE/?hc\_ref=SEARCH \\
Folha de SP              & https://www.facebook.com/folhadesp/                   \\
Revista Veja             & https://www.facebook.com/Veja/                        \\
O Antagonista            & https://www.facebook.com/oantagonista                 \\
Band News                & https://www.facebook.com/BandNews                     \\
Correio Braziliense      & https://www.facebook.com/correiobraziliense           \\
Jornal GGN               & https://www.facebook.com/JornalGGN/                   \\
MBL                      & https://www.facebook.com/mblivre/                     \\
Pragmatismo Político     & https://www.facebook.com/PragmatismoPolitico/         \\
Nexo                     & https://www.facebook.com/nexojornal/                 
\end{tabular}
\label{tab:pages}
\end{table*}

\end{appendix}

\bibliographystyle{plain}
\bibliography{references}

\end{document}